\documentclass[12pt]{article}

\usepackage{amsfonts,amsmath,amssymb,enumerate,epsfig,graphicx}

\numberwithin{equation}{section}

\newcommand{\ie}{{\it i.e.}\ }

\newcommand{\be}{\begin{equation}}
\newcommand{\ee}{\end{equation}}
\newcommand{\beu}{\begin{equation*}}
\newcommand{\eeu}{\end{equation*}}
\newcommand{\bea}{\begin{eqnarray}}
\newcommand{\eea}{\end{eqnarray}}
\newcommand{\beaa}{\begin{eqnarray*}}
\newcommand{\eeaa}{\end{eqnarray*}}
\newcommand{\bmx}{\begin{pmatrix}}
\newcommand{\emx}{\end{pmatrix}}

\def\cA{{\cal A}}                  
         \def\cE{{\cal E}}         \def\cF{{\cal F}}
                  \def\cI{{\cal I}}
                  
\def\cM{{\cal M}}                  
                  
\def\cS{{\cal S}}                  
\def\cV{{\cal V}}

\def\fa{{\mathfrak a}}

\def\fs{{\mathfrak s}}
\def\ft{{\mathfrak t}}

\newcommand{\CC}{{\mathbb C}}

\newcommand{\II}{{\mathbb I}}

\newcommand{\1}{\mbox{\hspace{.0em}1\hspace{-.24em}I}}

\newcommand{\mb}[1]{\quad\mbox{#1}\quad}

\newcommand{\half}{\frac{1}{2}}

\newcommand{\nonu}{\nonumber\\}
\newcommand{\End}{\mbox{End}}

\begin{document}

\pagestyle{empty}
\setcounter{page}{0}
\null
\vfill

\begin{center}

{\Huge \textsf{Direct computation of scattering matrices\\[1.7ex] 
for general quantum graphs}}

\vfill\vfill

{\large \textbf{V. Caudrelier}$^a$ and \textbf{E. Ragoucy$^{b}$ }}

\vfill
\emph{$^a$ Centre for Mathematical Science,
City University London,}

\emph{Northampton Square,
London,
EC1V 0HB,
United Kingdom.}
\vspace{2mm}\\
E-mail: v.caudrelier@city.ac.uk\\

\vspace{5mm}

\emph{$^b$ Laboratoire d'Annecy-le-Vieux de Physique Th{\'e}orique}

\emph{LAPTH, CNRS, UMR 5108, Universit{\'e} de Savoie}

\emph{9 chemin de Bellevue, B.P. 110, F-74941 Annecy-le-Vieux Cedex, 
France}\\

\vspace{2mm}

E-mail: ragoucy@lapp.in2p3.fr\\
\vfill

\begin{abstract}
We present a direct and simple method for the computation of the total 
scattering matrix of an arbitrary finite noncompact connected quantum graph 
given its metric structure and local scattering data at each vertex. The method 
is inspired by the formalism of Reflection-Transmission 
algebras and quantum field theory on graphs though the results hold independently 
of this formalism. 
It yields a simple and direct algebraic derivation of the formula for the total 
scattering and has a 
number of advantages compared to existing recursive methods. The case of loops 
(or tadpoles) is easily incorporated in our method. This provides 
an extension of recent similar results obtained in a completely different way in 
the context of abstract graph theory. It also allows us to discuss
briefly the inverse scattering problem in the presence of loops using an explicit 
example to show that the solution is not unique in general. On top of 
being conceptually very easy, the computational advantage of the method is 
illustrated on two examples of "three-dimensional" graphs (tetrahedron and 
cube) for which other methods are rather heavy or even impractical.

\end{abstract}
\end{center}
\vfill

\rightline{July 2009\qquad}
\rightline{LAPTH-1347/09\qquad}

\newpage
\pagestyle{plain}

\section{Introduction}

Excitement in the study of systems on quantum graphs has been revived recently 
as they provide models 
for the study of transport properties in quantum wires connected through 
junctions. It is largely motivated by the range of different physical applications 
that can be linked to such models, starting from condensed matter 
experiments or atomic wires up to chaos and neural networks, for 
reviews, see e.g. \cite{rev1,rev2}. 

A powerful formalism in this respect is that of quantum fields theory on 
graphs combined with bosonization techniques. One of the 
central objects in this approach is total scattering matrix of the graph and 
the knowledge of its analytic structure.
A number of results is already available in
\cite{BMS2,BBMS,BMS,BMS3} but essentially for star graphs.  
Results that apply to more general 
graphs can be found in \cite{KS3,KS2,KS1,Rag,KSS,KSS2,S}. 
However, all the different techniques presented 
in those papers imply the use of a recursive approach that becomes rather cumbersome when 
the graph is complicated.

The goal of this paper is to provide an efficient and simple techniques to 
compute this matrix for an arbitrary finite noncompact connected quantum graph 
knowing only its metric structure and local scattering data at each vertex. 
The point of view taken here is that the complete graph is obtained by assembling 
star graphs (single vertex graphs with a certain number of edges) which are 
well-understood. We obtain an explicit formula for the total scattering matrix.
It turns out that our results hold beyond the context of quantum field theory 
on graphs. Not only do they represent an extension of recent results \cite{KSS} 
to the case of graphs with loops but our method also provides a direct 
(as opposed to recursive \cite{Rag,S}) and simple algebraic derivation.

The paper is organised as follows. In section \ref{sec:general} we 
pesent our formalism to compute directly the scattering 
matrix associated to a general quantum graph without loop. Once the 
notation is settled, the calculation is very simple and effective. In 
the next section, we show how to extend the techniques to graphs with 
loops. Then, in section \ref{sec:platon}, we illustrate the techniques 
in computing the scattering matrix for graphs corresponding to Platonic 
solids, the cases of tetrahedron and cube being treated in great 
details. Finally, the last section is devoted to a short conclusion 
on possible applications.

\section{General setting and results\label{sec:general}}

We consider a finite noncompact graph with $N$ vertices that we label with 
$\alpha=1,\ldots,N$ and with internal and external edges. The graph is 
compact if it has no external edges. 
At each vertex $\alpha$ are attached possibly
several edges. One can endow the graph with a metric structure: 
the \textit{external edges} are associated to infinite half-lines and are 
connected to a unique vertex; the 
\textit{internal edges} are associated to intervals of finite length and 
connect two vertices, possibly not distinct. In the 
case where an internal edge connects the same vertex, we call it a loop 
(also called tadpole in the literature). Two edges are 
adjacent if they are connected by an internal edge. We consider a connected 
graph \ie a graph such that for any two vertices $\alpha$, 
$\beta$ there is a sequence $\{\alpha_1=\alpha,\alpha_2,\dots,\alpha_q=\beta\}$ 
of adjacent vertices. We define an 
orientation on the edges, and in the case of internal edges, 
$(\alpha\beta)$ will define an edge   
with orientation from vertex $\alpha$ to vertex $\beta$. By convention, 
external edges $(\alpha 0)$ are always oriented 
from the vertex to infinity. On each of these 
edges, we attach \textit{modes} (of fields living on the edge) 
$$
\fa_{j}^{\alpha\beta}(p)\quad j=1,\ldots,N_{\alpha\beta}\ ;\ 
\beta=0,1,\ldots,N\ ;\ \alpha=1,\ldots,N\ ; \ \alpha\neq \beta\,,
$$
$p$ being an orientation dependent parameter which has the interpretation 
of a momentum or a rapidity in applications to quantum fields on graphs and 
with the following conventions: 
\begin{itemize}
\item$\alpha=1,2,\ldots,N$ denotes the vertex to which 
the edge is attached; 
\item$\beta=0,1,2,\ldots,N$ denotes the vertex linked to $\alpha$ 
by the edge under consideration, with the convention that external 
edges corresponds to $\beta=0$;
\item  $j=1,\ldots,N_{\alpha\beta}$ 
numbers the different edges between $\alpha$ and $\beta$, 
$N_{\alpha\beta}$ being their total number. We set $N_{\alpha\beta}=0$ if 
$\alpha$ is not connected to $\beta$.
\end{itemize}
In this way the ordered 
triplet $(\alpha,\beta,j)$ uniquely defines all the oriented edges of the graph. 
Obviously, $(\alpha,\beta,j)$ and $(\beta,\alpha,j)$ define the same 
edge, but with a different orientation. Hence we have 
$N_{\alpha\beta}=N_{\beta\alpha}$. We will call internal mode (resp. 
external mode) a mode living on an internal edge (resp. external edge).

\subsection{General case without loops}\label{general}

For the time being, we assume $N_{\alpha\alpha}=0$ for all $\alpha=1,\dots,N$ 
\ie we do not consider loops. We will see later on that 
they are easily incorporated in our formalism. The modes are not independent 
but are related by two types of fundamental relations 
defining the scattering and propagation on the graph:
\begin{itemize}
\item Local scattering at vertex $\alpha$: Following the RT-algebra formalism 
(see e.g. \cite{RTalg1,RTalg2}), this reads 
\be
\label{local_comp}
\fa_{j}^{\alpha\beta}(p) = 
\sum_{\gamma=0}^N\sum_{k=1}^{N_{\alpha\gamma}} 
\fs^{\beta\gamma}_{\alpha;jk}(p)\, \fa_{k}^{\alpha\gamma}(-p)
\qquad\forall j=1,\ldots,N_{\alpha\beta}\ ;\ \forall\beta=0,1,\ldots,N
\ee
where $\fs^{\beta\gamma}_{\alpha;jk}(p)$ are the components of the local 
scattering matrix 
$S_\alpha(p)$ which satisfies $S_\alpha(p)S_\alpha(-p)=\1$.

\item Propagation on edge $(\alpha\beta j)$: As already mentionned, the 
edges $(\alpha\beta j)$ and $(\beta\alpha j)$ 
are identical (up to the orientation), so that the modes 
$\fa_{j}^{\alpha\beta}(p)$ and $\fa_{j}^{\beta\alpha}(p)$ are 
related. Denoting by $d^{\alpha\beta}_{j}=d^{\beta\alpha}_{j}$ the 
length of the edge, we have\footnote{The particular form of this relation 
comes from the fact that we have in mind applications to 
quantum fields which are Fourier transforms of the modes considered here, 
see \cite{Rag} for instance.}
\be
\label{propagation}
\fa_{j}^{\alpha\beta}(p) = 
\exp(-i\,d^{\alpha\beta}_{j}\,p)\,\fa_{j}^{\beta\alpha}(-p)\,.
\ee
\end{itemize}
The aim now is to obtain the scattering relations directly between the 
external modes \ie relations of the form
\be
\fa_{j}^{\alpha 0}(p)=
\sum_{\gamma=1}^N\sum_{k=1}^{N_{\gamma 0}} 
\fs^{\alpha\gamma}_{tot;jk}(p)\, \fa_{k}^{\gamma 0}(-p)
\qquad\forall j=1,\ldots,N_{\alpha 0}\ ;\ \forall\alpha=1,\ldots,N\,,
\ee
where $\fs^{\alpha\gamma}_{tot;jk}(p)$ are the components of the total 
scattering matrix for the graph, $S_{tot}(p)$.
This is most easily achieved by arranging the modes in vectors and using 
simple linear algebra. Denote 
$\cM_{r\times s}$ the vector space of $r\times s$ matrices over $\CC$. In 
particular, we identify $\cM_{n\times n}$ and $End(\CC^n)$. We denote 
$E^{r,s}_{i,j}$ the $r\times s$ matrix whose only nonzero entry is $1$ at 
position $(i,j)$. The set $\{E^{rs}_{ij}\}_{\stackrel{i=1,...,r;}{j=1,...,s}}$ 
is a basis of $\cM_{r\times s}$. We will drop the superscripts every time 
this does not cause confusion \ie each time the size of the matrix 
corresponds to the range of the indices. Similarly, we denote 
$\{e^n_j\}_{j=1,...,n}$ 
the canonical basis of $\CC^n$ and we will use a similar convention. Finally, 
we denote $\cF(p)$ the space of all (possibly generalized)
functions of $p\in\CC$, with the understanding that these functions can be 
operator-valued in quantum field applications (cf the modes). 
The following definitions illustrate our notations and conventions.
For a given vertex $\alpha$, we define different vectors:
\begin{itemize}
\item We collect the external modes attached to $\alpha$ in 
\bea
 A_{\alpha}(p)=\left(\begin{array}{c}
\fa_{1}^{\alpha0}(p)\\
\vdots\\
\fa_{N_{\alpha 0}}^{\alpha0}(p)
\end{array}\right)=\sum_{j=1}^{N_{\alpha0}} e_{j}\otimes 
\fa_{j}^{\alpha0}(p)\,\in \CC^{N_{\alpha 0}}\otimes \cF(p)
\eea

\item We collect the internal modes attached to $\alpha$ in
\bea 
B_{\alpha}(p)=\left(\begin{array}{c}
\fa_{1}^{\alpha 1}(p)\\
\vdots\\
\fa_{N_{\alpha 1}}^{\alpha 1}(p)\\
\fa_{1}^{\alpha 2}(p)\\
\vdots\\
\fa_{N_{\alpha2}}^{\alpha 2}(p)\\
\vdots\\
\vdots\\
\fa_{1}^{\alpha N}(p)\\
\vdots\\
\fa_{N_{\alpha N}}^{\alpha N}(p)
\end{array}\right)\,,
\eea
where only the modes with $N_{\alpha\beta}\neq 0$ appear.
For conciseness\footnote{The explicit, longer 
formula is 
$$
B_{\alpha}(p)=
\sum_{p=0}^{q_\alpha}\sum_{\beta=\beta_p+1}^{\beta_{p+1}-1}
\sum_{j=1}^{N_{\alpha\beta}}
e^{N-q_\alpha}_{\beta-p}\otimes e_{j}\otimes \fa_{j}^{\alpha\beta}(p)\,,
$$
where $\{\beta_1,\dots,\beta_{q_\alpha}\}$ are the labels $\beta$ such that 
$N_{\alpha\beta}=0$ and we have set $\beta_0=0$ and $\beta_{q_\alpha+1}=N+1$
for convenience. }, we write this as 
\bea
B_{\alpha}(p)=\sum_{\beta=1}^N\sum_{j=1}^{N_{\alpha\beta}}
e_{\beta}\otimes e_{j}\otimes \fa_{j}^{\alpha\beta}(p)\,
\in \CC^{\nu_{\alpha}}\otimes \cF(p)
\eea
where $\displaystyle \nu_\alpha=
\sum_{\beta=1}^N N_{\alpha\beta}$ is the number of internal edges 
attached to $\alpha$. This makes the following computations a lot more 
transparent but the reader should remember the actual content and 
size of each vector. 
 
\item Similarly, we collect all the modes attached to $\alpha$ in 
\bea
\cA_{\alpha}(p)=\sum_{\beta=0}^N\sum_{j=1}^{N_{\alpha\beta}}
e_{\beta+1}\otimes e_{j}\otimes \fa_{j}^{\alpha\beta}(p)\,\in 
\CC^{N_{\alpha}}\otimes \cF(p)
\eea
where $N_\alpha=N_{\alpha 0}+\nu_\alpha$
is the total number of edges attached to $\alpha$. This way, 
$\cA_\alpha$ is the concatenation of $A_\alpha$ and $B_\alpha$ with 
$A_\alpha$ "sitting on top".
\end{itemize}
With the same conventions, we introduce
\be
S_{\alpha}(p)= \sum_{\beta,\gamma=0}^N 
\sum_{j=1}^{N_{\alpha\beta}}\sum_{k=1}^{N_{\alpha\gamma}}
E_{\beta+1,\gamma+1}\otimes E_{jk}\otimes 
\fs^{\beta\gamma}_{\alpha;jk}(p)
\in \End(\CC^{N_\alpha})\otimes\cF(p)\,,
\ee
so the relations (\ref{local_comp}) read
\bea
\cA_\alpha(p)=
S_\alpha(p)\,\cA_\alpha(-p)\ ,\ \forall\alpha=1,\ldots,N\,.
\label{eq:Salpha}
\eea
The set of relations (\ref{eq:Salpha}) can be gathered into a single 
one:
\be
\cA(p) = S(p)\,\cA(-p) \mb{with}
\cA(p) =\sum_{\alpha=1}^N e_{\alpha}\otimes\cA_{\alpha}(p)
\mb{and} S(p) =\sum_{\alpha=1}^N E_{\alpha\alpha}\otimes S_{\alpha}(p)
\label{eq:ASA}
\ee
Remark that $\cA(p)\in \CC^{N_e+2N_i}\otimes\cF(p)$ where 
$\displaystyle N_e=\sum_{\alpha=1}^N N_{\alpha 0}$ is the total number 
of external edges and 
$\displaystyle N_i=\sum_{1\leq\alpha\leq\beta\leq N} N_{\alpha\beta}$ 
is the total number of internal edges.
Then, we introduce 
\be
B(p)=\sum_{\alpha=1}^N e_{\alpha}\otimes 
 B_{\alpha}(p)\,\in \CC^{2N_i}\otimes \cF(p)\,,
\ee
and
\be
\label{def_E(p)}
E(p) = \sum_{\alpha,\beta=1}^N 
\sum_{j=1}^{N_{\alpha\beta}} E_{\alpha,\beta}\otimes 
E_{\beta,\alpha}\otimes E_{jj}\otimes 
\exp(-i\,d^{\alpha\beta}_{j}\,p)\,\in End(\CC^{2N_i})\otimes \cF(p)\,,
\ee
so the relations (\ref{propagation}) read
\be
B(p)=E(p)B(-p)
\ee
It is easy to see that 
$$
E(p)\,E(-p) = \sum_{\alpha,\beta=1}^N\sum_{j=1}^{N_{\alpha\beta}} 
E_{\alpha,\alpha}\otimes 
E_{\beta,\beta}\otimes \II_{N_{\alpha\beta}}
$$
that acts as the identity matrix $\1_{2N_i}$.
The final step is to decompose the matrix $S(p)$ into four submatrices
related to external or internal edges:
\bea
\label{block1}
S^{(11)}(p) &=& \sum_{\alpha=1}^N
\sum_{j,k=1}^{N_{\alpha0}}E_{\alpha\alpha}\otimes E_{jk}\otimes 
\fs^{00}_{\alpha;jk}(p)\,\in End(\CC^{N_e})\otimes \cF(p)
\\
\label{block2}
S^{(12)}(p) &=& \sum_{\alpha,\gamma=1}^N
\sum_{j=1}^{N_{\alpha0}}\sum_{k=1}^{N_{\alpha\gamma}}
E_{\alpha\alpha}\otimes
E_{1,\gamma}\otimes E_{jk}\otimes \fs^{0\gamma}_{\alpha;jk}(p)\,\in 
\cM_{N_e\times 2N_i}\otimes \cF(p)
\\
\label{block3}
S^{(21)}(p) &=&\sum_{\alpha,\beta=1}^N 
\sum_{j=1}^{N_{\alpha\beta}}\sum_{k=1}^{N_{\alpha0}}
E_{\alpha\alpha}\otimes
E_{\beta, 1}\otimes E_{jk}\otimes \fs^{\beta0}_{\alpha;jk}(p)\,\in 
\cM_{ 2N_i\times N_e}\otimes \cF(p)
\\
\label{block4}
S^{(22)}(p) &=& \sum_{\alpha,\beta,\gamma=1}^N 
\sum_{j=1}^{N_{\alpha\beta}}\sum_{k=1}^{N_{\alpha\gamma}}
E_{\alpha\alpha}\otimes
E_{\beta,\gamma}\otimes E_{jk}\otimes \fs^{\beta\gamma}_{\alpha;jk}(p)\,
\in End(\CC^{2N_i})\otimes \cF(p)\,.\quad
\eea
Therefore, the set of all the relations we have becomes
\bea
A(p) &=&
S^{(11)}(p)\,A(-p)+ S^{(12)}(p)\,B(-p)
\\
B(p) &=&
S^{(21)}(p)\,A(-p)+ S^{(22)}(p)\,B(-p)\\
B(p)&=&E(p)B(-p)
\eea
Assuming that $E(p)-S^{(22)}(p)$ is invertible this yields the desired 
relations in the form
\be
A(p)=S_{tot}(p)A(-p)\,,
\ee
with
\be
\label{expression_Stot}
S_{tot}(p)=S^{(11)}(p)+S^{(12)}(p)\,\left[E(p)-S^{(22)}(p)\right]^{-1}
\,S^{(21)}(p)\,.
\ee
The internal modes can be expressed in terms of the external ones:
\be
B(p) = \left[E(-p)-S^{(22)}(-p)\right]^{-1}\, S^{(21)}(-p)\,A(p)
\label{expressionB}
\ee
These two formulas are the central result of this work. 
We note that in the course 
of our investigation, we discovered that 
the analog of the result (\ref{expression_Stot}) 
has been found in \cite{KSS} in the setting 
of abstract graph theory and using the formalism of Grassmann variables. 
However, the proof is based on the notion of generalized 
star product \cite{KS2} and requires a rather involved proof by induction on 
the size of the graph. Here, it is obtained directly by simple linear
algebra and ready to use for computations (either analytical or numerical).

\subsection{Discussion}

We have checked that our formula reproduces known results obtained by other 
methods for simple graphs (star-triangle, etc.) \cite{KS3,KS2,Rag}. 
In the following, we present in detail the computation for new graphs, 
especially in 3D, for which the previous methods are impractical. Our method 
presents
several advantages compared to previous ones. First, as just mentioned, it 
is computationally easier and one does not have to worry about the sequence 
of steps
used in iterative methods where one has to make sure that fusing two given 
vertices and then a third gives the same results a fusing the first and third 
and then the second (cf \cite{Rag}). The only task involved is the inversion of 
a matrix and there are well-known efficient methods both 
analytically and numerically.
Then, we have an explicit formula which shows the location of the poles of 
$S_{tot}$ (on top of the possible ones from the local matrices 
which are given data in our approach). They are solutions of 
\be
\label{quantization}
det (E(p)-S^{(22)}(p))=0\,.
\ee
This is important as these poles play a fundamental role in the computation 
of physical quantities like the conductance in quantum systems
 defined on graphs (see \cite{BMS,Rag}). Finally, for quantum systems on compact
graphs, \ie without external edges, the same equation provides the allowed 
modes on the graph. In this respect, (\ref{quantization}) 
is the generalization to an arbitrary compact quantum graph of the of the 
quantization equation
\be
e^{2ikL}=1\,,
\ee
for a particle in a box of length $L$. The matrix $S^{(22)}(p)$ accounts 
here for the one particle scattering occurring at the vertices. 
In the theory of integrable systems, this type of equations is sometimes 
called Bethe ansatz equations. 
However, here we emphasize that it is not related to such an ansatz. 
In condensed matter physics, the information given by this equation
together with the dispersion relation of the model provides the basis of 
band structure analysis.

\subsection{Properties}

To be consistent, our general formula should not depend on the numbering 
of the internal edges or vertices (internal permutation)
 and should transform appropriately under a permutation of the external 
modes (external permutation). Let $\Pi$ be an external permutation
  acting on $A(p)$ and $P$ an internal permutation acting on $B(p)$. 
It is easy to see that this induces the transformations
\bea
S^{(11)}(p)\to \Pi S^{(11)}(p)\Pi^{-1}\\
S^{(12)}(p)\to \Pi S^{(12)}(p)P^{-1}\\
S^{(21)}(p)\to P S^{(21)}(p)\Pi^{-1}\\
S^{(22)}(p)\to P S^{(22)}(p)P^{-1}\\
E(p)\to P E(p) P^{-1}\,,
\eea
producing $S_{tot}\to \Pi S_{tot}\Pi^{-1}$ as it should. Therefore, 
in examples or applications, one can always fix a convenient numbering 
of edges and vertices and 
work up to an external permutation.

In view of physical application, we must also be concerned with the 
properties of $S_{tot}$. 
We have seen already that $S_\alpha(p)S_\alpha(-p)=\1_{N_\alpha}$. This implies 
\be
S_{tot}(p)S_{tot}(-p)=\1_{N_e}\,.
\ee
To see this, note that the block matrix made of (\ref{block1})-(\ref{block4}) 
is related to $S(p)$ given in (\ref{eq:ASA}) by
\be
\cS(p)\equiv
\left(\begin{array}{c|c}
S^{(11)}(p) & S^{(12)}(p)\\
\hline
S^{(21)}(p) & S^{(22)}(p)
\end{array}\right)=P\,S(p)\,P^{-1}
\ee
where $P$ is the permutation matrix defined by 
\be
P\cA(p)=\left(\begin{array}{c}
A(p)\\
B(p)
\end{array}\right)\,.
\ee
Then by direct calculation and upon using $\cS(p)\cS(-p)=\1_{N_e+2N_i}$ 
and $E(p)E(-p)=\1_{2N_i}$ we get
\be
S_{tot}(p)S_{tot}(-p)=\1_{N_e}+S^{(12)}(p)\left(E(p)-S^{(22)}(p)\right)^{-1}
\,M\,\left(E(-p)-S^{(22)}(-p)\right)^{-1}S^{(21)}(-p)\,,
\ee
where
\bea
M&=&\1_{2N_i}-\left(E(p)-S^{(22)}(p)\right)\left(E(-p)-S^{(22)}(-p)\right)-
\left(E(p)
-S^{(22)}(p)\right)S^{(22)}(-p)\nonumber\\
&&-S^{(22)}(p)\left(E(-p)-S^{(22)}(-p)\right)-
S^{(22)}(p)S^{(22)}(-p)\nonumber\\
&=&0\,.
\eea
Now the local scattering matrices can be required to have additional 
properties, like unitarity. This is the case in particular if 
they arise from non-dissipative local boundary conditions emerging
from self-adjoint extensions of the free one-dimensional Hamiltonian 
(see e.g. \cite{KS3}). One then has unitarity 
\be
S_\alpha^\dagger(p)=S_\alpha^{-1}(p)\,.
\ee
Following the same type of argument as above, one finds that $S_{tot}$ 
is also unitary.

We finish this section by providing a few properties of $E(p)$. It is 
symmetric and we have already seen that $E(p)E(-p)=\1_{2N_i}$. In particular
$E(0)^2=\1_{2N_i}$ so its eigenvalues are $\pm 1$ and are equally degenerate. 
Also, $E(0)$ is a permutation matrix and 
$E(p)$ is a generalized permutation matrix (with coefficients of the type 
$e^{-ipd_j^{\alpha\beta}}$) which can be written as a product
of a permutation matrix and a diagonal matrix
\bea
E(p)=D(p)E(0)=E(0)D(p)\,,
\eea
where
\be
D(p)=\sum_{\alpha,\beta=1}^N 
\sum_{j=1}^{N_{\alpha\beta}} E_{\alpha,\alpha}\otimes 
E_{\beta,\beta}\otimes E_{jj}\otimes 
\exp(-i\,d^{\alpha\beta}_{j}\,p)\,,
\ee
with
\bea
D(p)D(q)=D(p+q)~~,~~p,q\in\CC\,.
\eea

\section{Including loops\label{sec:loops}}

The case of loops attached to single vertices can be treated with minor 
modifications in our formalism. Essentially, the idea is again to see a loop
attached to a given vertex $\alpha$ as arising from the gluing of two 
edges attached to this vertex. This will be most easily incorporated in the 
general formalism if we use the following trick for notations. Let 
$N_{\alpha\alpha}\neq 0$ be the number 
of loops attached to vertex $\alpha$. To each loop $j$, 
$j=1,\dots,N_{\alpha\alpha}$ correspond two modes\footnote{Again, 
the choice of numbering is 
 for convenience only and is irrelevant to the final results.} 
$a_{2j-1}^{\alpha\alpha}(p)$ and 
$a_{2j}^{\alpha\alpha}(p)$
which are 
related by
\be
a_{2j-1}^{\alpha\alpha}(p)=e^{-ipd_j^{\alpha\alpha}}a_{2j}^{\alpha\alpha}(-p)
~~,~~j=1,\dots,N_{\alpha\alpha}\,.
\ee
We collect these modes in two-component vectors
\be
\label{loopmodes}
\fa_j^{\alpha\alpha}(p)=\left(\begin{array}{c}
a_{2j-1}^{\alpha\alpha}(p)\\
a_{2j}^{\alpha\alpha}(p)
\end{array}\right)~~,~~j=1,\dots,N_{\alpha\alpha}\,.
\ee
We denote all the components of the local scattering matrix $S_\alpha(p)$ 
related to the loop modes by $s^{\alpha\beta}_{\alpha;jk}(p)$, $j=1,\dots,
2N_{\alpha\alpha}$, $k=1,\dots,N_{\alpha\beta}$; $s^{\beta\alpha}_{\alpha;jk}(p)$, 
$j=1,\dots,
N_{\alpha\beta}$, $k=1,\dots,2N_{\alpha\alpha}$; and 
$s^{\alpha\alpha}_{\alpha;jk}(p)$, $j,k=1,\dots,2N_{\alpha\alpha}$. 
Mimicking (\ref{loopmodes}), 
we then define, for $\alpha\neq \beta$,
\bea
\label{matrix_structure1}
\fs^{\alpha\beta}_{\alpha;jk}(p)=\left(\begin{array}{c}
s^{\alpha\beta}_{\alpha;2j-1,k}(p)\\
s^{\alpha\beta}_{\alpha;2j,k}(p)
\end{array}\right)~~,~~j=1,\dots,N_{\alpha\alpha},\,k=1,\dots,N_{\alpha\beta}\,,\\
\fs^{\beta\alpha}_{\alpha;jk}(p)=\left(\begin{array}{cc}
s^{\beta\alpha}_{\alpha;j,2k-1}(p)& s^{\beta\alpha}_{\alpha;j,2k}(p)
\end{array}\right)~~,~~j=1,\dots,N_{\alpha\beta},\,k=1,\dots,N_{\alpha\alpha}\,,
\eea
and also,
\bea
\label{matrix_structure2}
\fs^{\alpha\alpha}_{\alpha;jk}(p)=\left(\begin{array}{ll}
s^{\alpha\alpha}_{\alpha;2j-1,2k-1}(p)&s^{\alpha\alpha}_{\alpha;2j-1,2k}(p)\\
s^{\alpha\alpha}_{\alpha;2j,2k-1}(p)&s^{\alpha\alpha}_{\alpha;2j,2k}(p)
\end{array}\right)~~,~~j,k=1,\dots,N_{\alpha\alpha}\,.
\eea
Finally, we define 
\be
e_j^{\alpha\beta}(p)=\begin{cases}
e^{ipd_j^{\alpha\beta}}~~,~~\text{if}\, \beta\neq\alpha\,\text{and }\,
 N_{\alpha\beta}\neq 0\,,\\
e^{ipd_j^{\alpha\alpha}}\left(\begin{array}{cc}
0&1\\
1&0
\end{array}\right)~~,~~\text{if}\,\beta=\alpha\,\text{and }\, 
N_{\alpha\alpha}\neq 0\,.
\end{cases}
\ee
With all this, the relations defining scattering and propagation on 
the graph take the same form as before (cf (\ref{local_comp}) and 
(\ref{propagation})) 
\be
\fa_{j}^{\alpha\beta}(p) = 
\sum_{\gamma=0}^N\sum_{k=1}^{N_{\alpha\gamma}} 
\fs^{\beta\gamma}_{\alpha;jk}(p)\, \fa_{k}^{\alpha\gamma}(-p)
\qquad\forall j=1,\ldots,N_{\alpha\beta}\ ;\ \forall\beta=0,1,\ldots,N
\ee
and
\be
\fa_{j}^{\alpha\beta}(p) = 
e_j^{\alpha\beta}(-p)\,\fa_{j}^{\beta\alpha}(-p)\qquad\forall 
j=1,\ldots,N_{\alpha\beta}\ ;\ \forall\beta=0,1,\ldots,N\,.
\ee
Therefore, all the formalism and the results developed in section 
\ref{general} hold in the same form, provided one substitutes 
$e_j^{\alpha\beta}(-p)$
for $e^{-ipd_j^{\alpha\beta}}$ in (\ref{def_E(p)}). One should not be 
deceived by the apparent similarity of the results with or without loops. 
In general, 
the consequences of adding a loop in a given graph can be drastic.

However, as the formalism suggests, allowing for loops in graphs opens the 
possibility that two topologically completely different graphs 
can have exactly the same total scattering matrix. This is illustrated on 
the example below. In particular, this shows that the uniqueness of the 
inverse scattering problem, as discussed in \cite{KS1},
does not extend to the case of graphs with loops\footnote{Uniqueness is 
only guaranteed if one requires 
in addition that the number of vertices is maximal (cf Theorem 4.6 in \cite{KS1}).}.

We consider the two graphs depicted in Figure \ref{two_graphs} below. To 
illustrate our notations, we have displayed the modes involved in the 
construction, 
dropping the $p$-dependence for conciseness. They are topologically 
completely different, one being a triangle with one external edge attached 
to each 
vertex and the other being a single vertex star graph with three external 
edges and three loops attached to it. Note that for the triangle, we drop 
the unnecessary Latin subcripts since $N_{\alpha\beta}=1$ for all $\alpha=1,2,3$ 
and $\beta=0,1,2,3$, $\beta\neq \alpha$.
\begin{center}
\begin{figure}[ht]
\epsfig{file=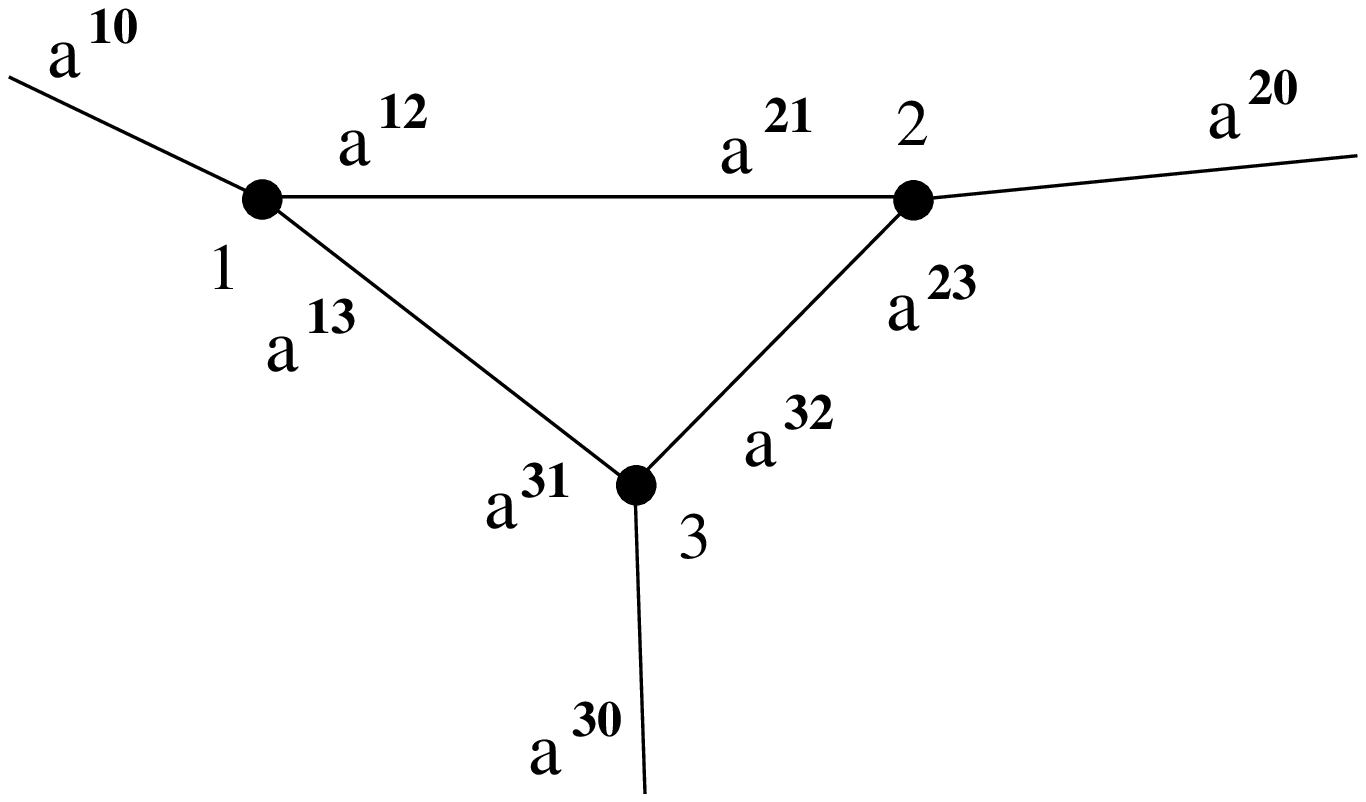,width=8cm}\hfill
\epsfig{file=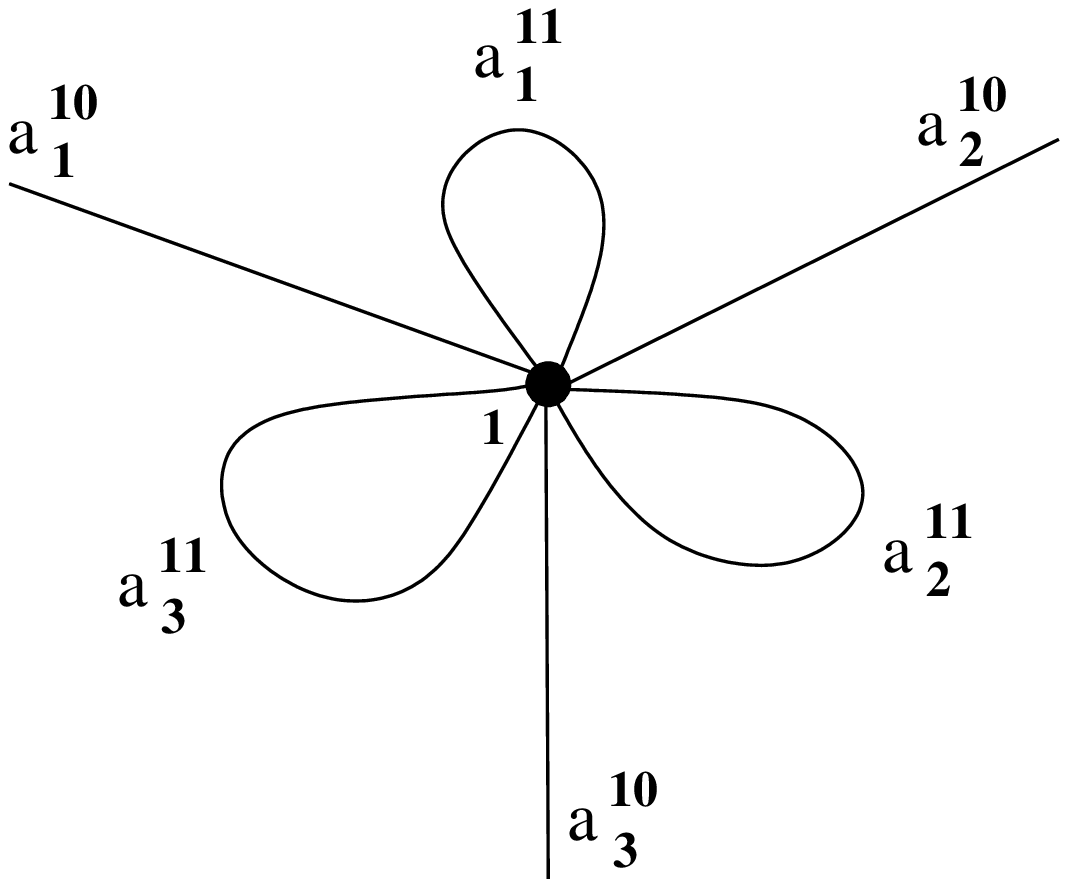,width=7cm}
\caption{\label{two_graphs}\textit{Two topologically different graphs with the same 
total scattering matrix. \newline
Left: triangle. Right: Star graph with loops.}}
\end{figure}
\end{center}

We assume that the scattering and propagation data 
is given as follows (we drop again the $p$ dependence for clarity),
\paragraph{For the triangle,}
\bea
S_1=\left(\begin{array}{ccc}
\fs_1^{00}&\fs_1^{02} &\fs_1^{03} \\
\fs_1^{20}&\fs_1^{22} &\fs_1^{23} \\
\fs_1^{30}&\fs_1^{32} &\fs_1^{33}
\end{array}\right)\,,~~
S_2=\left(\begin{array}{ccc}
\fs_2^{00}&\fs_2^{01} &\fs_2^{03} \\
\fs_2^{10}&\fs_2^{11} &\fs_2^{13} \\
\fs_2^{30}&\fs_2^{31} &\fs_2^{33}
\end{array}\right)\,,~~
S_3=\left(\begin{array}{ccc}
\fs_3^{00}&\fs_3^{01} &\fs_3^{02} \\
\fs_3^{10}&\fs_3^{11} &\fs_3^{12} \\
\fs_3^{20}&\fs_3^{21} &\fs_3^{22}
\end{array}\right)\,,\quad
\eea
giving the four blocks as defined in (\ref{block1})-(\ref{block4}) in the form
\bea
S^{(11)}=\left(\begin{array}{ccc}
\fs_1^{00}&0&0 \\
0&\fs_2^{00} &0 \\
0&0 &\fs_3^{00}
\end{array}\right)~~,~~S^{(22)}=\left(\begin{array}{cccccc}
\fs_1^{22} &\fs_1^{23}&0&0&0&0 \\
\fs_1^{32} &\fs_1^{33}&0&0&0&0\\
0&0&\fs_2^{11} &\fs_2^{13}&0&0 \\
0&0&\fs_2^{31} &\fs_2^{33}&0&0\\
0&0&0&0&\fs_3^{11} &\fs_3^{12} \\
0&0&0&0&\fs_3^{21} &\fs_3^{22}
\end{array}\right)\,,
\eea
\bea
S^{(12)}=\left(\begin{array}{cccccc}
\fs_1^{02} &\fs_1^{03}&0&0&0&0 \\
0&0&\fs_2^{01} &\fs_2^{03}&0&0\\
0&0&0&0&\fs_3^{01} &\fs_3^{02}
\end{array}\right)~~,~~
S^{(21)}=\left(\begin{array}{ccc}
\fs_1^{20} &0&0 \\
\fs_1^{30} &0&0\\
0&\fs_2^{10}&0 \\
0&\fs_2^{30}&0 \\
0&0&\fs_3^{10}\\
0&0&\fs_3^{20}
\end{array}\right)\,,\quad
\eea
and
\bea
E_t=\left(\begin{array}{cccccc}
0&0&e^{-ipd^{12}}&0&0&0\\
0&0&0&0&e^{-ipd^{13}}&0\\
e^{-ipd^{12}}&0&0&0&0&0\\
0&0&0&0&0&e^{-ipd^{23}}\\
0&e^{-ipd^{13}}&0&0&0&0\\
0&0&0&e^{-ipd^{23}}&0&0
\end{array}\right)\,.
\eea
\paragraph{For the star graph,}
\bea
T=\left(\begin{array}{cccccc}
\ft_{1;11}^{00}&0 &0&\ft_{1;11}^{01} &\ft_{1;12}^{01}&\ft_{1;13}^{01}  \\
0&\ft_{1;22}^{00}&0&\ft_{1;21}^{01}&\ft_{1;22}^{01}&\ft_{1;23}^{01} \\
0&0&\ft_{1;33}^{00}&\ft_{1;31}^{01} &\ft_{1;32}^{01}&\ft_{1;33}^{01}\\
\ft_{1;11}^{10}&\ft_{1;12}^{10}&\ft_{1;13}^{10}&\ft_{1;11}^{11}
&\ft_{1;12}^{11}&\ft_{1;13}^{11}\\ 
\ft_{1;21}^{10}&\ft_{1;22}^{10}&\ft_{1;23}^{10}&\ft_{1;21}^{11}
&\ft_{1;22}^{11}&\ft_{1;23}^{11}\\
\ft_{1;31}^{10}&\ft_{1;32}^{10}&\ft_{1;33}^{10}&\ft_{1;31}^{11}
&\ft_{1;32}^{11}&\ft_{1;33}^{11}
\end{array}\right)\equiv
\left(\begin{array}{c|c}
T^{(11)}&T^{(12)}\\
\hline
T^{(21)}&T^{(22)}
\end{array}\right)\,,
\eea
and 
\bea
E_s=\left(\begin{array}{cccccc}
0&e^{-ipd_1^{11}}&0&0&0&0\\
e^{-ipd_1^{11}}&0&0&0&0&0\\
0&0&0&e^{-ipd_2^{11}}&0&0\\
0&0&e^{-ipd_2^{11}}&0&0&0\\
0&0&0&0&0&e^{-ipd_3^{11}}\\
0&0&0&0&e^{-ipd_3^{11}}&0
\end{array}\right)\,.
\eea

The lengths of the internal edges of the triangle are related to the lengths 
of the loop in the star graph by
\bea
d^{12}=d_1^{11}~~,~~d^{23}=d_3^{11}~~,~~d^{23}=d_2^{11}\,,
\eea
and the following relations for the scattering data hold, showing in particular 
the matrix structure defined in (\ref{matrix_structure1})-
(\ref{matrix_structure2}) in the case of loops, 
\bea
&&\ft_{1;11}^{00}=\fs_1^{00},\,\ft_{1;22}^{00}=\fs_2^{00},\,
\ft_{1;33}^{00}=\fs_3^{00}\,,\\
&&\ft_{1;11}^{01}=\left(\begin{array}{cc}
\fs_1^{02} &0
\end{array}\right),\,\ft_{1;12}^{01}=\left(\begin{array}{cc}
0 &0
\end{array}\right),\,\ft_{1;13}^{01}=\left(\begin{array}{cc}
\fs_1^{03} &0
\end{array}\right)\,,\\
&&\ft_{1;21}^{01}=\left(\begin{array}{cc}
0&\fs_2^{01} 
\end{array}\right),\,\ft_{1;22}^{01}=\left(\begin{array}{cc}
\fs_2^{03} &0
\end{array}\right),\,\ft_{1;23}^{01}=\left(\begin{array}{cc}
0 &0
\end{array}\right)\,,\\
&&\ft_{1;31}^{01}=\left(\begin{array}{cc}
0 &0
\end{array}\right),\,\ft_{1;32}^{01}=\left(\begin{array}{cc}
0&\fs_3^{02} 
\end{array}\right),\,\ft_{1;33}^{01}=\left(\begin{array}{cc}
0&\fs_3^{01} 
\end{array}\right)\,,\\
&& \ft_{1;11}^{10}=\left(\begin{array}{c}
\fs_1^{20} \\
0
\end{array}\right),\,\ft_{1;12}^{10}=\left(\begin{array}{c}
0\\
\fs_2^{10}
\end{array}\right),\,\ft_{1;13}^{10}=\left(\begin{array}{c}
0\\
0
\end{array}\right)\,,\\
&&\ft_{1;21}^{10}=\left(\begin{array}{c}
0\\
0
\end{array}\right),\,\ft_{1;22}^{10}=\left(\begin{array}{c}
\fs_2^{30} \\
0
\end{array}\right),\,\ft_{1;23}^{10}=\left(\begin{array}{c}
0\\
\fs_3^{20}
\end{array}\right)\,,\\
&&\ft_{1;31}^{10}=\left(\begin{array}{c}
\fs_1^{30} \\
0
\end{array}\right),\,\ft_{1;32}^{10}=\left(\begin{array}{c}
0\\
0
\end{array}\right),\,\ft_{1;33}^{10}=\left(\begin{array}{c}
0\\
\fs_3^{10}
\end{array}\right)\,,\\
&&\ft_{1;11}^{11}=\left(\begin{array}{cc}
\fs_1^{22}&0\\
0&\fs_2^{11}
\end{array}\right),\,\ft_{1;12}^{11}=\left(\begin{array}{cc}
0&0\\
\fs_2^{13}&0
\end{array}\right),\,\ft_{1;13}^{11}=\left(\begin{array}{cc}
\fs_1^{23}&0\\
0&0
\end{array}\right)\,,\\ 
&&\ft_{1;21}^{11}=\left(\begin{array}{cc}
0&\fs_2^{31}\\
0&0
\end{array}\right),\,\ft_{1;22}^{11}=\left(\begin{array}{cc}
\fs_2^{33}&0\\
0&\fs_3^{22}
\end{array}\right),\,\ft_{1;23}^{11}=\left(\begin{array}{cc}
0&0\\
0&\fs_3^{21}
\end{array}\right)\,,\\
&&\ft_{1;31}^{11}=\left(\begin{array}{cc}
\fs_1^{32}&0\\
0&0
\end{array}\right),\,\ft_{1;32}^{11}=\left(\begin{array}{cc}
0&0\\
0&\fs_3^{12}
\end{array}\right),\,\ft_{1;33}^{11}=\left(\begin{array}{cc}
\fs_1^{33}&0\\
0&\fs_3^{11}
\end{array}\right)\,.
\eea
The fact that these two graphs give rise to the same total scattering matrix 
follows from the fact that their scattering data are related 
by an internal permutation
\be
P=\left(\begin{array}{cccccc}
1&0&0&0&0&0\\
0&0&1&0&0&0\\
0&0&0&1&0&0\\
0&0&0&0&0&1\\
0&1&0&0&0&0\\
0&0&0&0&1&0
\end{array}\right)\,,
\ee
such that 
\bea
P\,S^{(22)}\,P^{-1}=T^{(22)}~~,~~P\,E_tP^{-1}=E_s~~,~~S^{(12)}\,P^{-1}=T^{(12)}
~~,~~P\,S^{(21)}=T^{(21)}\,.
\eea
Then,
\bea
S^{triangle}_{tot}&=&S^{(11)}+S^{(12)}\left(E_t-S^{(22)}\right)^{-1}S^{(21)}
\nonumber\\
&=&S^{(11)}+S^{(12)}\,P^{-1}\left(P\,E_t\,P^{-1}-P\,S^{(22)}\,
P^{-1}\right)^{-1}P\,S^{(21)}\nonumber\\
&=&T^{(11)}+T^{(12)}\left(E_s-T^{(22)}\right)^{-1}T^{(21)}\nonumber\\
&=&S^{star}_{tot}\,.
\eea

\section{Platonic solids\label{sec:platon}}

In this section, we illustrate the freedom on numbering and the use of 
formula (\ref{expression_Stot}) on the convex regular polyhedra known as 
Platonic solids (tetrahedron, cube,
octahedron, dodecahedron, icosahedron) \cite{platon}. 
Once the scattering matrix is 
known, physical quantities associated to the graph can easily be 
computed, such as the conductance, using the formalism developped in 
\cite{BMS2}. The calculation essentially relies on the pole structure and 
the general techniques have been explicited in \cite{Rag}.

We carry out explicit calculations  
in the case of the tetrahedron and the cube. 
This choice is primarily motivated by aesthetic and academic criteria rather 
than any particular practical 
application. It also shows the computational advantage of our method over 
recursive ones on rather involved graphs. 
More precisely, we consider graphs whose 
internal edges and vertices correspond to Platonic solids and for which exactly 
one external edge is attached to each vertex. This corresponds to 
$N_{\alpha\beta}=1$, $\alpha=1,\dots,N$, $\beta=0,\dots,N$, $\alpha\neq \beta$.
 Note that the condition of regularity yields $d_1^{\alpha\beta}\equiv d$ 
for all $\alpha,\beta=1,\dots,N$. Also, all the vertices are connected to the 
same number of vertices so
$\nu_\alpha\equiv \nu$ is the same for all $\alpha=1,\dots,N$. $N$ is even for 
all those graphs. Finally, from the general theory of graph colouring,
see e.g. \cite{color}, 
it is known that 
we can assign a label (or colour) $a\in\{1,\dots,\nu\}$ to the edges connected 
to the same vertex in a way compatible with the graph \ie, in colour terms, 
such that no two edges connected to the same vertex have the same colour and 
each edge can only have one colour. This allows us to define functions
$n_\alpha$, $\alpha\in\{1,\dots,N\}$ from $\{1,\dots,\nu\}$ to $\{1,\dots,N\}$ 
such that 
$n_\alpha(a)=\beta$ if and only if $\beta$ is connected to $\alpha$ by the edge 
labelled $a$. We use the convention $a=0$ for external edges and set 
$n_\alpha(0)=0$ for all $\alpha\in\{1,\dots,N\}$. By construction, we have the 
following properties
\bea
n_\alpha(a)=\beta \Leftrightarrow n_\beta(a)=\alpha~~,~~n_\alpha(a)
=n_\beta(a)\Leftrightarrow \alpha=\beta~~,~~n_\alpha(a)=n_\alpha(b)
\Leftrightarrow a=b\,.
\eea
In view of formula (\ref{expression_Stot}), the main object of interest 
is $E(p)-S^{(22)}(p)$ which we seek to invert. 
With our notations, we get
\bea
E(p)=e^{-ipd}\sum_{\alpha=1}^N\sum_{a=1}^\nu E_{\alpha,n_\alpha(a)}\otimes 
E_{aa}~~,~~S^{(22)}(p)=\sum_{\alpha=1}^N\sum_{a,b=1}^\nu E_{\alpha,\alpha}
\otimes E_{ab}\otimes \fs_\alpha^{ab}(p)\,,
\eea
where the local matrices read
\be
S_\alpha(p)=\sum_{a,b=0}^\nu E_{a+1,b+1}\otimes \fs_\alpha^{ab}(p)~~,~~
\alpha=1,\dots,N \,.
\ee
For later convenience, we define a reduced scattering matrix containing only 
the information about scattering on the internal edges
\be
S^{red}_\alpha(p)=\sum_{a,b=1}^\nu E_{a,b}\otimes \fs_\alpha^{ab}(p)~~,~~
\alpha=1,\dots,N \,.
\ee
Let us also define
\bea
\cE_a=\sum_{\alpha=1}^N E_{\alpha,n_\alpha(a)}\,.
\eea
Then $\displaystyle E(p)=e^{-ipd}\sum_{a=1}^\nu \cE_a \otimes E_{aa}$ 
and from the general properties of $E$ (or by direct calculation) we find
\bea
\cE_a=\cE_a^t=\cE_a^{-1}~~,~~a=1,\dots,\nu\,.
\eea
Therefore $\cE_a$ is diagonalizable with eigenvalues $\pm 1$ each 
degenerate $\frac{N}{2}$ times and with eigenvectors $v_\alpha^\epsilon=
\frac{1}{\sqrt{2}}(e_\alpha+\epsilon\, e_{n_\alpha(a)})$, $\epsilon=\pm 1$, 
$\alpha< n_\alpha(a)$, forming an orthonormal basis.

\subsection{Tetrahedron}
\begin{center}
\begin{figure}[ht]
\hspace{3cm} \epsfig{file=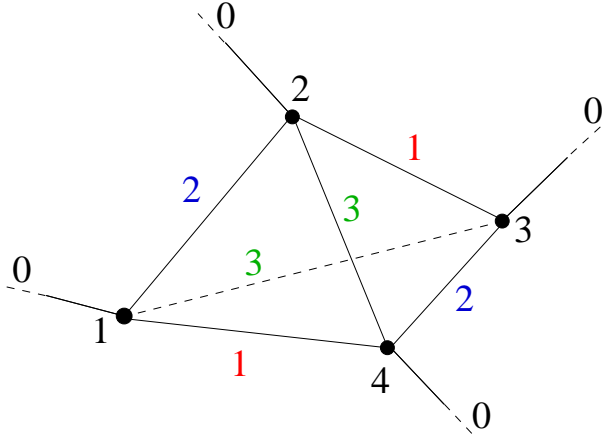,width=8cm}
\caption{\label{tetraedre}\textit{Tetrahedron with an example of 
numbering.}}
\end{figure}
\end{center}

For the tetrahedron, $N=4$, $\nu=3$ and the matrices $\cE_a$ enjoy 
the additional property
\be
\label{commute}
\cE_a\cE_b=\cE_b\cE_a~~,~~\forall ~a,b=1,2,3\,,
\ee
due to the fact that 
\be
\forall \beta\in\{1,2,3,4\},\ \forall a,b\in\{1,2,3\}~~,~~n_{n_\beta(a)}(b)
=n_{n_\beta(b)}(a)\,.
\ee
This can be seen to hold by direct inspection on figure \ref{tetraedre} and 
holds also for other inequivalent numberings.
From (\ref{commute}), they can be diagonalized simultaneously. As already 
explained, to fix ideas we can fix a numbering without loss of generality since
we work up to permutations. In the present case, changing the edges and or 
vertices numbering amounts to interchanging the $\cE_a$'s.
From the figure we obtain
\bea
\cE_1=\left(\begin{array}{cccc}
0&0&0&1\\
0&0&1&0\\
0&1&0&0\\
1&0&0&0
\end{array} \right)~~,~~\cE_2=\left(\begin{array}{cccc}
0&1&0&0\\
1&0&0&0\\
0&0&0&1\\
0&0&1&0
\end{array} \right)~~,~~\cE_3=\left(\begin{array}{cccc}
0&0&1&0\\
0&0&0&1\\
1&0&0&0\\
0&1&0&0
\end{array} \right)\,,
\eea
and a diagonalizing matrix is 
\be
T=\frac{1}{2}\left(\begin{array}{cccc}
-1&1&-1&1\\
1&-1&-1&1\\
-1&-1&1&1\\
1&1&1&1
\end{array} \right)=T^{-1}=T^t\,.
\ee
So far, we haven't taken advantage of the geometry and its symmetries. The 
scattering can still be different from vertex to vertex (as labelled by the 
index $\alpha$ on the local matrices) and at a given vertex, the scattering 
from edge $a$ to edge $b$ needs not be the same as the scattering from edge
 $a$ to edge $c$ say. Clearly, this does not respect the natural symmetry of 
the underlying graph. One can impose that the local scattering matrices
 be the same for all vertices \ie $S_\alpha(p)\equiv S(p)$ and in particular 
$S^{red}(p)\equiv S^{red}(p)$ for all $\alpha=1,2,3,4$. This already greatly 
 simplifies the problem of inversion. Let $\tau=T\otimes\1_3$ and 
$D_a=T\cE_a T^{-1}$ then
 \bea
 \tau(E(p)-S^{(22)}(p))\tau^{-1}=e^{-ipd}\sum_{a=1}^3D_a\otimes 
E_{aa}-\1_4\otimes S^{red}(p)
 \eea
 The matrix on the right-hand side is a block diagonal matrix made of four 
$3\times 3$ blocks essentially determined by $S^{red}$
 \be
 \tau(E(p)-S^{(22)}(p))\tau^{-1}=\sum_{\alpha=1}^4E_{\alpha\alpha}\otimes 
\left(e^{-ipd}I_\alpha-S^{red}(p)\right)\,,
 \ee
with 
\bea
I_1=\left(\begin{array}{ccc}
-1&0&0\\
0&-1&0\\
0&0&1
\end{array} \right),~I_2=\left(\begin{array}{ccc}
1&0&0\\
0&-1&0\\
0&0&-1
\end{array} \right),~I_3=\left(\begin{array}{ccc}
-1&0&0\\
0&1&0\\
0&0&-1
\end{array} \right),~I_4=\1_3\,.\quad
\eea
Thus, the problem is reduced to inverting $3\times 3$ matrices. In particular, 
the poles of $S_{tot}$ are solutions of
\be
det\left(e^{-ipd}I_\alpha-S^{red}(p)\right)=0~~,~~\alpha=1,2,3,4\,.
\ee
We now turn to the explicit calculation of $S_{tot}$ in the case where the 
vertices are described by scale invariant local matrices (independent of $p$) 
capturing universal features of scattering. In our case, each local matrix is 
the same $4\times 4$ scale invariant matrix whose explicit form has been 
classified in \cite{BMS2}. 
Note also that we can take further advantage of the symmetries of the underlying 
geometry here by imposing for instance that the scattering be invariant 
under a rotation of $\pi/3$ around the axis passing through a vertex and the centre of 
the opposite face. Physically, this means 
that an incoming particle from the external edge of a vertex has the same probability of 
being transmitted to any one of the internal edges attached to this
vertex. Mathematically, this amounts to requiring that $S$ satisfies
\be
\left(\begin{array}{cc}
1&0\\
0&J
\end{array}\right)S\left(\begin{array}{cc}
1&0\\
0&J^{-1}
\end{array}\right)=S\,,
\ee 
where 
\be
J=\left(\begin{array}{ccc}
0&1&0\\
0&0&1\\
1&0&0
\end{array} \right)~~,~~J^3=\1_3\,.
\ee
Putting everything together, we find two possible local scattering matrices
\bea
\label{2cases}
S_1=\left(
\begin{array}{cccc}
 -\frac{1}{2} & \frac{1}{2} & \frac{1}{2} & \frac{1}{2} \\
 \frac{1}{2} & -\frac{1}{2} & \frac{1}{2} & \frac{1}{2} \\
 \frac{1}{2} & \frac{1}{2} & -\frac{1}{2} & \frac{1}{2}\\
 \frac{1}{2} & \frac{1}{2} & \frac{1}{2} & -\frac{1}{2}
\end{array}
\right)~~,~~
S_2=\left(
\begin{array}{cccc}
 -\frac{1}{2} & \frac{1}{2} & \frac{1}{2} & \frac{1}{2} \\
 \frac{1}{2} & \frac{5}{6} & -\frac{1}{6} & -\frac{1}{6} \\
 \frac{1}{2} & -\frac{1}{6} & \frac{5}{6} & -\frac{1}{6} \\
 \frac{1}{2} & -\frac{1}{6} & -\frac{1}{6} & \frac{5}{6}
\end{array}
\right)\,.
\eea
In the first case, we compute the $S_{tot}(p)$ as
\be
S^1_{tot}(p)=\frac{1}{G_1(p)}
 \left(-2 \left(e^{-3ipd}+e^{-ipd}-1\right) \1_4+ e^{-ipd} (e^{-ipd}+1)A \right)\,,
 \ee
 where $G_1(p)=(2e^{-2ipd}+e^{-ipd}+1)(2e^{-ipd}-1)$ and
 \be
 A=
\left(
\begin{array}{cccc}
0 & 1 & 1 & 1 \\
 1 & 0 & 1 & 1 \\
 1 & 1 & 0 & 1 \\
1 & 1 & 1 & 0
\end{array}\,.
\right)
\ee
The poles of this matrix are given by
\be
e^{-ipd} = x \mb{with} x\in\Big\{ 
\half,\,\frac{-1+i\sqrt7}{4},\,\frac{-1-i\sqrt7}{4}\Big\}\,.
\ee 

In the second case, we obtain
\be
S^2_{tot}(p)=\frac{1}{G_2(p)}
\left(-2 \left(-6 e^{-3ipd}+4 e^{-2ipd}+10 e^{-ipd}-6\right) \1_4+ 3 
e^{-ipd}(e^{-ipd}-1)A \right)\,,
\ee
where $G_2(p)=(6e^{-2ipd}-e^{-ipd}-3)(2e^{-ipd}-1)$, leading to the 
poles
\be
e^{-ipd} = x \mb{with} x\in\Big\{ 
\half,\,\frac{1+\sqrt{73}}{12},\,\frac{1-\sqrt{73}}{12}\Big\}\,.
\ee 

\subsection{cube}
For the cube, $N=8$ and $\nu=3$ and the matrices $\cE_a$ also commute. 
So one can perform the same analysis as before.
\begin{center}
\begin{figure}[ht]
\hspace{3cm} \epsfig{file=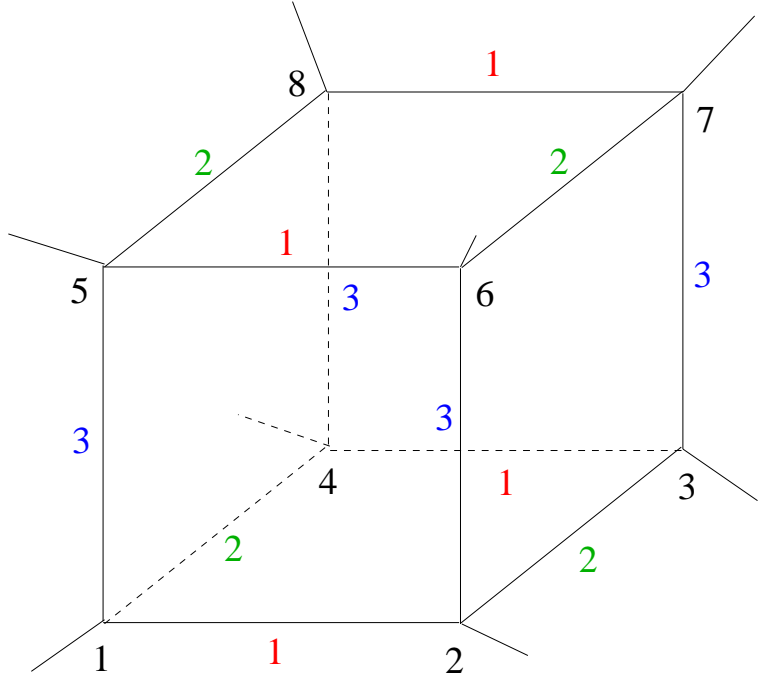,width=10cm}
\caption{\textit{\label{cube}Cube with an example of numbering.}}
\end{figure}
\end{center}
Based on figure \ref{cube}, we get explicitly
\bea
\cE_1=\left(
\begin{array}{cccccccc}
 0 & 1 & 0 & 0 & 0 & 0 & 0 & 0 \\
 1 & 0 & 0 & 0 & 0 & 0 & 0 & 0 \\
 0 & 0 & 0 & 1 & 0 & 0 & 0 & 0 \\
 0 & 0 & 1 & 0 & 0 & 0 & 0 & 0 \\
 0 & 0 & 0 & 0 & 0 & 1 & 0 & 0 \\
 0 & 0 & 0 & 0 & 1 & 0 & 0 & 0 \\
 0 & 0 & 0 & 0 & 0 & 0 & 0 & 1 \\
 0 & 0 & 0 & 0 & 0 & 0 & 1 & 0
\end{array}
\right)~~,~~\cE_2=\left(
\begin{array}{cccccccc}
 0 & 0 & 0 & 1 & 0 & 0 & 0 & 0 \\
 0 & 0 & 1 & 0 & 0 & 0 & 0 & 0 \\
 0 & 1 & 0 & 0 & 0 & 0 & 0 & 0 \\
 1 & 0 & 0 & 0 & 0 & 0 & 0 & 0 \\
 0 & 0 & 0 & 0 & 0 & 0 & 0 & 1 \\
 0 & 0 & 0 & 0 & 0 & 0 & 1 & 0 \\
 0 & 0 & 0 & 0 & 0 & 1 & 0 & 0 \\
 0 & 0 & 0 & 0 & 1 & 0 & 0 & 0
\end{array}
\right)\,,
\eea
\bea
\cE_3=\left(
\begin{array}{cccccccc}
 0 & 0 & 0 & 0 & 1 & 0 & 0 & 0 \\
 0 & 0 & 0 & 0 & 0 & 1 & 0 & 0 \\
 0 & 0 & 0 & 0 & 0 & 0 & 1 & 0 \\
 0 & 0 & 0 & 0 & 0 & 0 & 0 & 1 \\
 1 & 0 & 0 & 0 & 0 & 0 & 0 & 0 \\
 0 & 1 & 0 & 0 & 0 & 0 & 0 & 0 \\
 0 & 0 & 1 & 0 & 0 & 0 & 0 & 0 \\
 0 & 0 & 0 & 1 & 0 & 0 & 0 & 0
\end{array}
\right)\,,
\eea
and a diagonalizing matrix is 
\be
V=\frac{1}{2\sqrt{2}}\left(
\begin{array}{cccccccc}
 1 & -1 & 1 & -1 & -1 & 1 & -1 & 1 \\
 1 & 1 & -1 & -1 & -1 & -1 & 1 & 1 \\
 -1 & 1 & 1 & -1 & 1 & -1 & -1 & 1 \\
 -1 & -1 & -1 & -1 & 1 & 1 & 1 & 1 \\
 -1 & 1 & -1 & 1 & -1 & 1 & -1 & 1 \\
 -1 & -1 & 1 & 1 & -1 & -1 & 1 & 1 \\
 1 & -1 & -1 & 1 & 1 & -1 & -1 & 1 \\
 1 & 1 & 1 & 1 & 1 & 1 & 1 & 1
\end{array}\right)=(V^{-1})^t\,.
\ee
Again assuming that the local scattering matrices are the same at all vertices, we get 
\bea
 \cV(E(p)-S^{(22)}(p))\cV^{-1}=e^{-ipd}\sum_{a=1}^3\Delta_a\otimes 
E_{aa}-\1_8\otimes S^{red}(p)\,,
 \eea
 where $\Delta_a=V\,\cE_a\, V^{-1}$ and $\cV=V\otimes \1_3$. 
This is a block diagonal matrix and the problem is reduced to inverting $3\times 3$
 matrices,
 \bea
 (E(p)-S^{(22)}(p))^{-1}=\cV^{-1}\left[\sum_{\alpha=1}^8E_{\alpha\alpha}\otimes 
\left(e^{-ipd}\cI_\alpha - S^{red}(p)\right)^{-1}\right]\,\cV\,,
 \eea
where 
\bea
\cI_1=\1_3,~\cI_2=\left(\begin{array}{ccc}
-1&0&0\\
0&1&0\\
0&0&1
\end{array} \right),~\cI_3=\left(\begin{array}{ccc}
1&0&0\\
0&-1&0\\
0&0&1
\end{array} \right),~\cI_4=\left(\begin{array}{ccc}
-1&0&0\\
0&-1&0\\
0&0&1
\end{array} \right),~\\
\cI_5=\left(\begin{array}{ccc}
1&0&0\\
0&1&0\\
0&0&-1
\end{array} \right),~\cI_6=\left(\begin{array}{ccc}
-1&0&0\\
0&1&0\\
0&0&-1
\end{array} \right),~\cI_7=\left(\begin{array}{ccc}
1&0&0\\
0&-1&0\\
0&0&1
\end{array} \right),~\cI_8=-\1_3\,.\quad
\eea
We turn to the explicit computation of the total scattering matrix in the two cases 
(\ref{2cases})
describing scale and rotation invariant local scattering at the vertices. In both cases,
 we find the following structures for $S_{tot}$: it is a linear
combination of matrices in the abelian group generated by the $\cE$'s with 
coefficients being polynomials in $e^{-ipd}$. For $j=1,2$,
\bea
S^j_{tot}(p) &=& a_0^j(p)\1_8+a_1^j(p)\cE_1+a_2^j(p)\cE_2+a_3^j(p)
\cE_3+a_4^j(p)\cE_1\cE_2
+a_5^j(p)\cE_1\cE_3+a_6^j(p)\cE_2\cE_3
\nonu
&& +a_7^j(p)\cE_1\cE_2\cE_3\,.
\eea
In the first case, we find
\bea
a_0^1(p)&=&\frac{8+e^{-ipd}-8 e^{-2ipd}-5 e^{-3ipd}-40 e^{-4ipd}+4 e^{-5ipd}-32 
e^{-6ipd}}{4 \left(-1+e^{-2ipd}+8 e^{-4ipd}+16 e^{-6ipd}\right)}
\,,\\
a_1^1(p)&=&\frac{-5 e^{-ipd}+e^{-3ipd}-20 e^{-5ipd}}{4 \left(-1+e^{-2ipd}+8 e^{-4ipd}
+16 e^{-6ipd}\right)}
\,,\\
a_2^1(p)&=&\frac{3 e^{-ipd}}{4-16 e^{-2ipd}}
\,,\\
a_3^1(p)&=&\frac{3 e^{-ipd}}{4-16 e^{-2ipd}}
\,,\\
a_4^1(p)&=&-\frac{e^{-ipd} \left(1-9 e^{-ipd}+2 e^{-2ipd}\right)}{4 \left(-1+e^{-ipd}
+2 e^{-2ipd}-4 e^{-3ipd}+8 e^{-4ipd}\right)}
\,,\\
a_5^1(p)&=&-\frac{e^{-ipd} \left(1-9 e^{-ipd}+2 e^{-2ipd}\right)}{4 \left(-1+e^{-ipd}
+2 e^{-2ipd}-4 e^{-3ipd}+8 e^{-4ipd}\right)}
\,,\\
a_6^1(p)&=&\frac{e^{-ipd} \left(1+9 e^{-ipd}+2 e^{-2ipd}\right)}{4 \left(-1-e^{-ipd}
+2 e^{-2ipd}+4 e^{-3ipd}+8 e^{-4ipd}\right)}
\,,\\
a_7^1(p)&=&-\frac{e^{-ipd}+19 e^{-3ipd}+4 e^{-5ipd}}{4 \left(-1+e^{-2ipd}+8 e^{-4ipd}
+16 e^{-6ipd}\right)}\,.
\eea
The poles of the scattering matrix can be then computed. They are 
given by
\be
e^{-ipd} = x \mb{with} x\in\Big\{ 
\pm\half,\,\pm\frac{1+i\sqrt7}{4},\,\pm\frac{1-i\sqrt7}{4}\Big\}\,.
\ee 
In the second case, we find
\bea
a_0^2(p)&=&\frac{72+9 e^{-ipd}-440 e^{-2ipd}-45 e^{-3ipd}+728 e^{-4ipd}+36 e^{-5ipd}
-288 e^{-6ipd}}{4 \left(-9+73 e^{-2ipd}-184 e^{-4ipd}+144 e^{-6ipd}\right)}
\,,\qquad\\
a_1^2(p)&=&-\frac{3 e^{-ipd} \left(15-67 e^{-2ipd}+60 e^{-4ipd}\right)}{4 \left(-9+73 
e^{-2ipd}-184 e^{-4ipd}+144 e^{-6ipd}\right)}
\,,\\
a_2^2(p)&=&\frac{3 e^{-ipd}}{4-16 e^{-2ipd}}
\,,\\
a_3^2(p)&=&\frac{3 e^{-ipd}}{4-16 e^{-2ipd}}
\,,\\
a_4^2(p)&=&-\frac{3 e^{-ipd} \left(-1+3 e^{-ipd}+2 e^{-2ipd}\right)}{4 
\left(3-e^{-ipd}-18 e^{-2ipd}+4 e^{-3ipd}+24 e^{-4ipd}\right)}
\,,\\
a_5^2(p)&=&-\frac{3 e^{-ipd} \left(-1+3 e^{-ipd}+2 e^{-2ipd}\right)}{4 
\left(3-e^{-ipd}-18 e^{-2ipd}+4 e^{-3ipd}+24 e^{-4ipd}\right)}
\,,\\
a_6^2(p)&=&\frac{3 e^{-ipd} \left(-1-3 e^{-ipd}+2 e^{-2ipd}\right)}{4 
\left(3+e^{-ipd}-18 e^{-2ipd}-4 e^{-3ipd}+24 e^{-4ipd}\right)}
\,,\\
a_7^2(p)&=&\frac{3 e^{-ipd} \left(3-7 e^{-2ipd}+12 e^{-4ipd}\right)}{4 
\left(9-73 e^{-2ipd}+184 e^{-4ipd}-144 e^{-6ipd}\right)}\,.
\eea
The poles of the scattering matrix are 
given by
\be
e^{-ipd} = x \mb{with} x\in\Big\{ 
\pm\half,\,\pm\frac{1+\sqrt{73}}{12},\,\pm\frac{1-\sqrt{73}}{12}\Big\}\,.
\ee 

\section{Conclusion and outlooks}

In this paper, we have presented a direct method for the computation of the total 
scattering matrix of an arbitrary finite noncompact connected graph 
given its topology, metric structure and local scattering data at each vertex. The 
method uses the formalism of quantum modes as our initial motivation 
was the study of quantum fields on graphs. This resulted in a simple and direct 
algebraic derivation of formula (\ref{expression_Stot}). We have also
shown that the case of loops is easily incorporated in our method. This has been 
illustrated with an explicit example whose purpose was also to point out
that the inverse scattering problem on graphs does not have a unique solution in 
general for graphs with loops. 

We want to stress that in the present paper, the \textit{modes} as we called them, 
appear more as convenient labels than true quantum field theoretic 
objects. This has two consequences. First, our results are ready to use for 
applications in quantum field theory on graphs by simply promoting the modes to 
generators of the RT-algebra \cite{RTalg1}. Second, it means that our results hold in 
complete generality for abstract graphs with or without loops. In this 
respect, the present results provide an extension of the results in \cite{KSS} to 
the case of loops\footnote{The more general case where relations 
(\ref{propagation}) are replaced by a general invertible connecting matrix is easily 
implemented in our context.}.

Finally, this paper lays the ground to applications to transport problems on arbitrary 
graphs in the spirit of the studies performed in e.g. 
\cite{BMS2,BMS,BBMS,BMS3}. Indeed, it provides one with the central ingredient which 
is the total scattering matrix together with its pole 
structure. Once this structure is known, the calculation of physical 
data such as the
conductance between external edges is rather direct, see e.g. 
\cite{Rag}: we will return to these questions in the near future.


\begin{thebibliography}{10}
\bibitem{rev1} P. Exner, \textit{Leaky Quantum Graphs: A Review},
Proc. Symp. Pure Math. \textbf{77} (2008) 523 and 
\texttt{arXiv:0710.5903}.

\bibitem{rev2} P. Kuchment,
\textit{Quantum graphs: an introduction and a brief survey}, 
Proc. Symp. Pure Math. \textbf{77} (2008) 291 and 
\texttt{arXiv:0802.3442}.

\bibitem{BMS2} B. Bellazzini, M. Mintchev, P. Sorba, \textsl{Bosonization and 
Scale Invariance on Quantum Wires},
J. Phys. \textbf{A40} (2007) 2485 and \texttt{hep-th/0611090}.

\bibitem{BBMS} B. Bellazzini, M. Burrello, M. Mintchev, P. Sorba, \textsl{Quantum 
Field Theory on Star Graphs}, Proc. Symp. Pure Math. 
\textbf{77} (2008) 639 and \texttt{arXiv:0801.2852}.

\bibitem{BMS} B. Bellazzini, M. Mintchev, P. Sorba, \textsl{Boundary Bound State 
Effects in Quantum Wires}, 
\texttt{arXiv:0810.3101}.

\bibitem{BMS3} B. Bellazzini, M. Mintchev, P. Sorba, \textsl{Quantum wire junctions 
breaking time reversal invariance}, \texttt{arXiv:0907.4221}.

\bibitem{KS3}  V. Kostrykin, R. Schrader, \textsl{Kirchhoff's Rule for Quantum Wires}, 
J. Phys. \textbf{A32} (1999) 595-630
and \texttt{math-ph/9806013}.

\bibitem{KS2} V. Kostrykin and R. Schrader, \textsl{The generalized star product and 
the factorization of scattering matrices on
graphs}, J. Math. Phys. {\bf42} (2001) 1563 and \texttt{math-ph/0008022}.

\bibitem{KS1}  V. Kostrykin, R. Schrader,
\textsl{The inverse scattering problem for metric graphs and the traveling salesman 
problem}, \texttt{math-ph/0603010}.

\bibitem{Rag} E. Ragoucy, \textsl{Quantum field theory on quantum graphs and 
application to their conductance}, J. Phys. \textbf{A42} (2009) 295205 and 
\texttt{arXiv:0901.2431}.

\bibitem{KSS} Sh. Khachatryan, R. Schrader, A. Sedrakyan, \textsl{Grassmann-Gaussian 
integrals and generalized star products},
\texttt{arXiv:0904.2683}

\bibitem{KSS2} Sh. Khachatryan, A. Sedrakyan, P. Sorba, 
\textsl{Network Models: Action formulation},
\texttt{arXiv:0904.2688}

\bibitem{S} R. Schrader, \textsl{Finite propagation speed and causal free quantum 
fields on networks}, \texttt{arXiv:0907.1522}.

\bibitem{RTalg1}
M.~Mintchev, E.~Ragoucy and P.~Sorba,
\textit{Scattering in the presence of a reflecting and transmitting 
impurity},
Phys. Lett.  {\bf B547} (2002) 313 and
\texttt{hep-th/0209052};\\
\textit{Reflection-Transmission algebras},
J. Phys. {\bf A36} (2003) 10407 and
\texttt{hep-th/0303187};
 
\bibitem{RTalg2}
V.~Caudrelier, M.~Mintchev, E.~Ragoucy and P.~Sorba,
\textit{Reflection-Transmission quantum Yang-Baxter equations},
J. Phys. {\bf A38} (2005) 3431 and
\texttt{hep-th/0412159}.

\bibitem{platon} Pythagoras, \textsl{autos ephe}, $\sim$ 500 B.C.\\
Theaetetus, Book X of \textsl{Euclid's Elements}, 300 B.C.

\bibitem{color} C. Shannon, 
\textsl{A theorem on coloring the lines of a network}, 
J. Math. Phys. \textbf{28} (1949) 148Ð151;\\
V.G. Vizing, \textsl{On an estimate of the chromatic class of a p-graph}, Diskret. 
Analiz. 3 (1964) 25Ð30;\\
\textsl{Critical graphs with given chromatic class}, Metody Diskret. Analiz. 
\textbf{5} (1965) 9Ð17;\\
P. Erd\"os and R. Wilson,
\textsl{Note on the chromatic index of almost all graphs}, 
J. Combinatorial Theory, \textbf{B23} (1977) 255Ð257.

\end{thebibliography}
\end{document}